\documentclass[aps,showpacs,twocolumn]{revtex4}%
\usepackage{amsfonts}%
\usepackage{amsmath}%
\usepackage{amssymb}%
\usepackage{graphicx}

\begin{document}
\title{Asymmetric Properties of Heat Conduction in a One-Dimensional
Frenkel-Kontorova Model}
\author{Bambi Hu$^{1,2}$, Lei Yang$^{1}$, and Yong Zhang$^{1,\ast}$%
\footnotetext{$^{\ast}$Author to whom correspondence should be
address. Electronic address: yzhang@phys.hkbu.edu.hk}}
\affiliation{$^{1}$Department of Physics, Centre for Nonlinear
Studies, and The Beijing-Hong Kong-Singapore Joint Centre for
Nonlinear and Complex Systems (Hong Kong), Hong Kong Baptist
University, Kowloon Tong, Hong Kong, China\\$^{2}$Department of
Physics, University of Houston, Houston, Texas 77204-5005}


\begin{abstract}
In this Letter, we show numerically that the rectifying effect of
heat flux in a one-dimensional two-segment Frenkel-Kontorova chain
demonstrated in recent literature is merely available under the
limit of the weak coupling between the two constituent segments.
Surprisingly, the rectifying effect will be reversed when the
properties of the interface and the system size change. The two
types of asymmetric heat conduction are dominated by different
mechanisms, which are all induced by the nonlinearity. We further
discuss the possibility of the experimental realization of thermal
diode or rectifier devices.

\end{abstract}

\maketitle

Understanding of the phonon transport behavior under nonequilibrium
stationary states is much less than the electron in charge
conduction. It is a key distinctness between them that the former is
related to nonequilibrium statistic mechanics, while the latter is
basically the transport behavior under thermal equilibrium states.
Even for the phonon thermal conduction in the simplest solid model,
such as the Fermi-Pasta-Ulam $\beta$ model, in low dimensional
cases, it is still an open question to get full microscopic dynamics
description of macroscopic thermal transport behavior, i.e. the
Fourier law\cite{Bonetto2000,Lepri2003,Livi2003,ChaosFocus}. On the
other hand, the band theory predicts the unidirectional transport of
the electrons in heterogeneity semiconductor junctions, which have
very useful applications in modern electronics. This property is
basically attributed to the facts that the electrons obey Fermi
statistics and only the valence electrons take part in the transport
process. In contrast, the phonons obey Bose-Einstein statistics and
all take part in thermal transport. In this respect, a phonon,
viewed as a collective exciton in crystal lattices, is not expected
to exhibit intrinsic asymmetric transport properties, unless the
heat conduction of material dramatically changes when the heat baths
are exchanged.

Recently, however, the asymmetric heat conduction in one-dimensional
inhomogeneous chains with nonlinear on-site potentials has been
reported via computer
simulations\cite{Terraneo2002,Baowen2004,Hu2005,Baowen2005}. They
proposed some mechanisms which allow the heat flux in one direction
while the systems act almost like an insulator when the heat baths
are exchanged. These mechanisms are based on a common idea that
phonon bands of the different segments of the chain change from
overlap to separation when the heat baths are exchanged. The phonon
band shift controlled by temperature is attributed to the different
contribution of nonlinearity to the effective phonon spectra at
different temperatures. The effective phonon spectrum can be
obtained quantitatively by self-consistent phonon
method\cite{Dauxois1993} and qualitatively by linearization
treatment of Hamiltonians\cite{Baowen2004}. On the other hand, the
numerical results in previous work\cite{Baowen2004,Hu2005} show that
the rectifying effect decreases as the system size and the coupling
of the interface increase. Nevertheless, the picture of
overlap/separation of phonon bands, which treats each of the
different segments of the chain independently, is totally
independent of the system size and the coupling of the interface.
Thus, this picture fails to explain numerical results on the
decreasing of the rectifying effect. Surprisingly, as shown by our
numerical results in this paper, the rectifying effect will be
reversed when these parameters further increase. So, more in-depth
understanding is needed to assess possible applications of this
counterintuitive phenomenon discovered by computer simulations.

In this Letter, we implement extensive numerical studies about a
one-dimensional two-segment Frenkel-Kontorova (FK) chain that is
exactly the same as the model studied by Ref.\cite{Baowen2004}. We
find that the rectifying effect demonstrated in
Ref.\cite{Baowen2004} will be reversed when the properties of the
interface and the system size change. A simple physical picture was
proposed to understand the reversal of the rectifying effect, and a
new but very intuitive mechanism for the new type of asymmetric heat
transport was addressed. We further argue that the experimental
realization is a hard task, because the asymmetric heat conduction
dramatically depends on the properties of the interface and the
system size.

We consider a one-dimensional chain consisting of two segments of
$N/2$ particles. The two segments are connected by a linear spring
of constant $k_{int}$. The Hamiltonian of the whole system is
\begin{equation}
H=H_{A}+H_{B}+\frac{1}{2}k_{int}(x_{N/2+1}-x_{N/2}-a)^{2}, \label{WH}%
\end{equation}
where $H_{A}$ and $H_{B}$ are the Hamiltonian of the left segment (A
segment) and the right one (B segment), respectively, and both are
the FK chains described by
\begin{equation}
H=\underset{i=1}{\overset{N/2}{\sum}}\frac{p_{i}^{2}}{2m}+\frac{1}{2}%
k(x_{i+1}-x_{i}-a)^{2}-\frac{V}{(2\pi)^{2}}\cos2\pi x_{i}. \label{SH}%
\end{equation}
We set the mass of the particles and the lattice constant $m=a=1$,
and fix $V_{A}=5,V_{B}=1,k_{A}=1$ and $k_{B}=0.2$. The main
adjustable parameters in this paper are $k_{int\text{ }}$and $N$.

In our numerical simulations we use fixed boundary conditions and
the chain is coupled, at the two ends, with heat baths at
temperatures $T_{+}$ and $T_{-}$ respectively. We take $T_{+}=0.105$
and $T_{-}=0.035$ through our paper. We use Nos\'{e}-Hoover heat
baths and integrate the differential equations of motion by using
the fourth-order Runge-Kutta algorithm as described in
\cite{Press1992}. We compute the temperature profile inside the
system, i.e.,
the local temperature at site $i$ defined as $T_{i}=m\langle\overset{.}{x_{i}%
}^{2}\rangle$, where $\langle\rangle$ stands for temporal average,
the local heat flux
$j_{i}=k\langle\overset{.}{x}_{i}(x_{i}-x_{i-1})\rangle
$\cite{Lepri2003} and $J=Nj$ the total heat flux. The simulations
are performed long enough to allow the system to reach a steady
state where the local heat flux is constant along the chain. We
denote the absolute value of total heat flux from left to right as
$J_{+}$ when the bath at higher temperature $T_{+}$ is at the left
end of the chain and as $J_{-}$ when the
baths are exchanged.%

\begin{figure}
[ptb]
\begin{center}
\includegraphics[
height=3.4074in, width=2.5503in
]%
{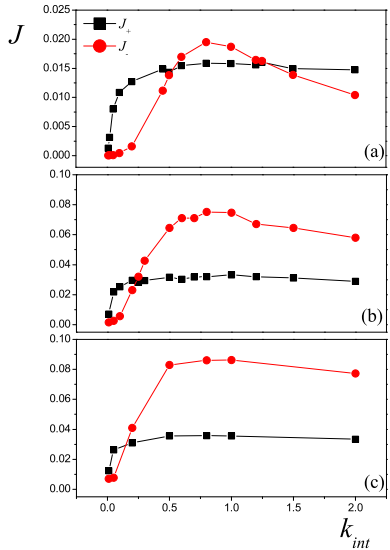}%
\caption{Total heat current $J_{\pm}$ versus interface harmonic
coupling
strength $k_{int}$ for (a) $N=100$; (b) $N=1000$; (c) $N=2000$.}%
\end{center}
\end{figure}

Figure 1 shows $k_{int}$ dependence of $J_{+}$ and $J_{-}$ in
different system size $N=100$, $1000$, $2000$. As $k_{int}$
increases there exist a crossover from the situation with
$J_{+}>J_{-}$ to that with $J_{+}<J_{-}$, that is, the rectifying
effect is reversed. In the situation $J_{+}>J_{-}$, our numerical
results are in agreement with those reported in
Ref.\cite{Baowen2004} at the same parameters and the prediction by
the effective-phonon analysis (as shown by the Fig.4 in the
Ref.\cite{Baowen2004}). When the heat flux flows from A to B there
is a big overlap between the phonon bands of A and B, while there is
almost a separation between them when the baths are exchanged. Only
phonons with the frequencies within the overlap range have
contributions to heat flux. So $J_{+}>J_{-}$ and $J_{-}$ is near
zero. Moreover, the overlapping bandwidth between A and B is almost
unchanged when the two segments are very weakly coupled. So in this
case the $J_{\pm}$ are only determined by the transmission of the
phonons crossing the interface, i.e., $k_{int}$. One can expect that
$J_{\pm}$ increase in the same way as $k_{int}$ increases. This is
confirmed by the numerical result that $j_{\pm}\sim k_{int}^{2}$
with small $k_{int}$ and $N$ (as shown by the Fig.2(a) in
Ref.\cite{Baowen2004}). In fact, this relationship is universal when
one consider the transmission of phonons crossing the interface
between two kind of material that are weakly joined by a harmonic
spring\cite{Patton2001,Hu20052}. In the situation $J_{+}<J_{-}$, the
results are strikingly different. $J_{\pm}$ have sharp increases
compared to the former case, and almost independent of the $k_{int}$
as shown in Fig.1(b) and Fig.1(c). These results are disagreement
with the above prediction based on the effective-phonon analysis,
and strongly suggest that phonon bands of A and B mix up. It is
obvious that the band mixing is due to the interaction among phonons
induced by the nonlinearity in the chain. In fact, the
effective-phonon analysis in Ref\cite{Baowen2004} treated each of
the two segments of the chain independently, and it could be based
on the condition $k_{int}\rightarrow0$. Consequently, the
prediction, i.e. $J_{+}>J_{-}$, is well agreement with the numerical
results for very small $k_{int}$ (a typical value in simulations is
$0.05$). As $k_{int}$ increases the two-segment chain behaves as a
whole. The phonons of the chain in this situation become different
from the ones of A and B due to the band mixing. Therefore, the
individual effective-phonon analysis to each of A and B fails to
explain the new type of asymmetric heat conduction.

There is the difference between curves of the $J_{+}$ and $J_{-}$ in
Fig.1(a). $J_{+}$ reaches saturation as $k_{int}$ increases. $J_{-}$
reaches its maximum value at the middle $k_{int}$ value with
$(k_{A}+k_{B})/2=0.6$, but decreases as $k_{int}$ further increases.
Specially, $J_{+}$ is larger than $J_{-}$ again when $k_{int}$
larger than the value $k_{int}\simeq k_{A}+k_{B}=1.2$ as shown in
Fig.1(a). These phenomena can be understood as follow. The role of
$k_{int}$ changes to an impurity in the two-segment chain when $k_{int}%
\gg(k_{A}+k_{B})/2$. The impurity just scatters the high-frequency
phonons, and does not affect the low-frequency ones. We speculate
that, for the mixing phonon spectrum, the high-frequency phonons
have main contributions to $J_{-}$ and the low-frequency ones to
$J_{+}$. As a result, $J_{-}$ drops due to the impurity scattering,
while $J_{+}$ is almost unchanged as $k_{int}$ increases. The effect
of the impurity on the heat conduction reduces with $N$ increasing.
So the reduction of $J_{-}$ decrease as $N$ increase as shown by
Fig.1. It is clear seen that both $J_{+}$ and $J_{-}$ almost
saturate as $k_{int}$ increases as shown in Fig.1(c). This means
that the heat conduction behavior of the chain is almost independent
of $k_{int}$ when the two segments are well coupled. Note also that
the crossover occurs at $k_{int}\approx0.5$, $0.24$, $0.15$ for
$N=100$, $1000$, $2000$. This result strongly implies that the
crossover will also appears as $N$ increases under certain fixed
$k_{int}$. It
is confirmed by Fig.2.%

\begin{figure}
[ptb]
\begin{center}
\includegraphics[
height=2.1283in, width=3.4091in
]%
{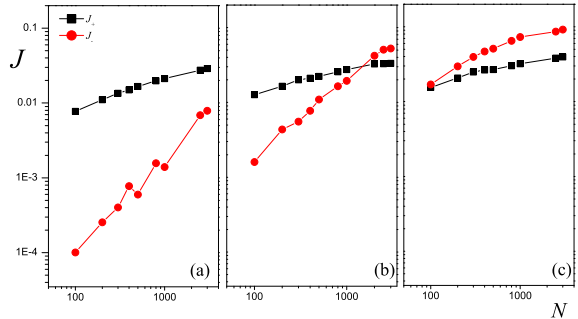}%
\caption{Total heat current $J_{\pm}$ versus the system size $N$ in
the
log-log plot for (a) $k_{int}=0.05$; (b) $k_{int}=0.2$; (c) $k_{int}=0.6$.}%
\end{center}
\end{figure}

Figure 2 depicts the $N$ dependence of $J_{+}$ and $J_{-}$ at
different $k_{int}$. For a typical weak coupling value,
$k_{int}=0.05$, the crossover does not occur, i.e. $J_{+}>J_{-}$, up
to $N=3000$. It is clear, however, seen that the increasing of
$J_{-}$ is more faster than $J_{+}$ as shown in Fig.2(a). This
implies that the crossover will occur at a larger $N$. In Fig.2(b),
the crossover appears at $N\approx1500$ for $k_{int}=0.2$. In
Fig.2(c), the two-segment chain exhibits $J_{+}<J_{-}$ for the whole
regime of $N$ used for simulations. This result can also be
understood by the role of $k_{int}$ in the limit case
$N\rightarrow\infty$. Strictly speaking, there is not a geometric
boundary within a one-dimensional system in thermodynamic limit,
unless the coupling of the two segments is exact zero. For any
finite $k_{int}$, the interface of two segments behaves like a
impurity in the limit $N\rightarrow\infty$, and the effect of the
impurity on heat conduction of the whole system reduces with $N$
increasing, as stated above. This suggest that the two-segment chain
will exhibit the behavior of $J_{+}<J_{-}$ in the thermodynamical
limit.

Up to now, it is clear seen that there are two types of asymmetric
heat conduction in the two-segment FK model (\ref{WH}), and the
transition between them will occur by varying the $k_{int}$ and $N$.
When the two segments are very weakly coupled, $J_{+}$ is larger
than $J_{-}$; while $J_{-}$ is larger than $J_{+}$ when the two
segments are well coupled or the chain is long enough. Phonon band
shift and phonon mixing play important roles in the former case and
the latter case, respectively. It is interesting that these two
contrary behaviors are all attributed to the nonlinearity.%

\begin{figure}
[ptb]
\begin{center}
\includegraphics[
height=3.1548in, width=2.7319in
]%
{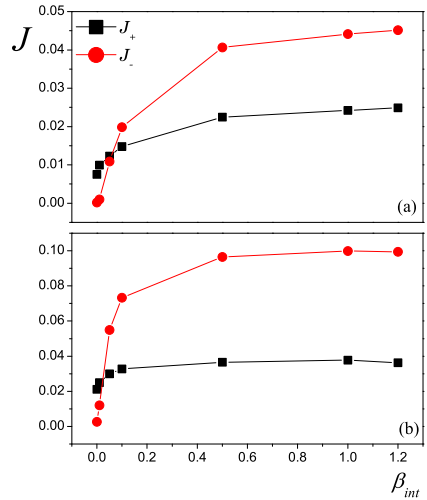}%
\caption{Total heat current $J_{\pm}$ versus interface anharmonic
coupling
strength $\beta_{int}$ for (a) $N=100$; (b) $N=1000$.}%
\end{center}
\end{figure}

In order to further verify the role of nonlinearity in the
asymmetric heat conduction in model (\ref{WH}), we add a quartic
nonlinear coupling to the
interface in model (\ref{WH}), as described by%

\begin{multline}
H=H_{A}+H_{B}+\frac{1}{2}k_{int}(x_{N/2+1}-x_{N/2}-a)^{2}\\
+\frac{1}{4}\beta_{int}(x_{N/2+1}-x_{N/2}-a)^{4}. \label{NF2}%
\end{multline}
\newline Figure 3 plots the $J_{\pm}$ versus $\beta_{int}$ for $N=100$ and
$N=1000$ at a typical weak coupling parameter $k_{int}=0.05$. By
increasing the $\beta_{int}$, one can also see the similar crossover
as shown in Fig.1 and Fig.2. The crossover occurs at
$\beta_{int}\simeq0.06$ and $\beta _{int}\simeq0.02$ for $N=100$ and
$1000$, respectively. These results confirm that the phonon mixing
due to the nonlinearity leads to the crossover from the
case $J_{+}>J_{-}$ to the one $J_{+}<J_{-}$, even if the nonlinearity is very small.%

\begin{figure}
[ptb]
\begin{center}
\includegraphics[
height=2.2355in, width=2.7051in
]%
{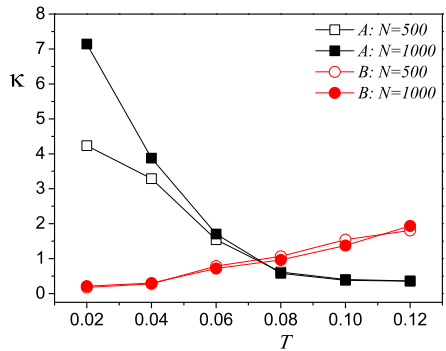}%
\caption{Heat conductivity $\kappa$ versus temperature $T_{0}$ for
segment A
and segment B, respectively. The temperatures of heat baths are $T_{\pm}%
=T_{0}(1\pm\delta)$. $T_{0}=0.02$, $0.04$, $0.06$, $0.08$, $0.1$,
$0.12$, and
$\delta=0.1$. $\kappa=J/(2\delta T_{0})$.}%
\end{center}
\end{figure}

A problem immediately arises from the above analysis: why is $J_{-}$
larger than $J_{+}$ when the phonon bands of A and B mix up? In the
thermodynamical limit, it can be expected that the heat conduction
of this two-segment chain is independent of the properties of
interface and the system size and determined by the heat transport
properties of A and B. In Fig.4, we plot the temperature dependence
of the heat conductivity $\kappa$ for A and B, respectively. As the
temperature change from $0.02$ to $0.12$, $\kappa_{A}$ decreases,
and $\kappa_{B}$ increase. Thus, both A and B have the higher heat
conductivity when A is contacted with $T_{-}=0.035$ and B with
$T_{+}=0.105$. When the baths are exchanged, both A and B have the
lower heat conductivity. It is very intuitive that $J_{-}$ is larger
than $J_{+}$.

In summary, we find the reversal of the rectifying effect in the
one-dimensional two-segment FK model via computer simulations. We
distinguish two different (even contrary) effects of the
nonlinearity on the heat conduction behaviors. For the weak coupling
of the two segments, the phonon band shift due to the nonlinearity
leads to $J_{+}>J_{-}$. In an other case, the nonlinearity causes
the mixing of phonon bands and leads to $J_{+}<J_{-}$. These two
types of asymmetric heat conduction can transit to each other by
varying the properties of the interface and the system size.

The asymmetric heat conduction discovered by computer simulations
opens new possibilities to design a thermal rectifier or thermal
diode in theory\cite{Terraneo2002,Baowen2004}. However, as shown by
our numerical results, the function of such a thermal device will
dramatically depend on the properties of the interface. To exactly
control the interface properties is very difficult in the
lab\cite{Swartz1989}, and this difficulty really leads to the poor
reproducibility of the experimental measurement of the thermal
boundary resistance. As a result, the function of the thermal
devices based on the principles stated here will be unpredictable
due to the unknown interface conditions, such as the weak nonlinear
atom-atom interaction. Thus, this is a hard task to make a thermal
rectifier or thermal diode in the lab.

YZ is indebted to Prof. Hong Zhao for very helpful discussions. We
would also like to thank Dr. D. He and members of the Centre for
Nonlinear Studies for useful discussions. This work was supported in
part by grants from the Hong Kong Research Grants Council (RGC) and
the Hong Kong Baptist University Faculty Research Grant (FRG).

\end{document}